\documentstyle[amsbsy]{article}
\makeatletter

          \@addtoreset{equation}{section}
       \makeatother

\newcommand{\ymh}{Yang-Mills-Higgs Lagrangian$\;$}
\newcommand{\p}{\partial}

\newcommand{\be}{\begin{equation}}
\newcommand{\ee}{\end{equation}}
\newcommand{\bea}{\begin{eqnarray}}
\newcommand{\eea}{\end{eqnarray}}
\newcommand{\bd}{\begin{displaymath}}
\newcommand{\ed}{\end{displaymath}}
\newcommand{\ds}{\displaystyle}
\renewcommand{\a}{&\!\!\!\!\!&\!\!\!\!\!}
\newcommand{\lab}[1]{\label{#1}\\}
\begin{document}
\title{%
%\vskip 1cm
\bf BRST invariant formulation of spontaneously broken gauge theory 
in generalized differential geometry\\
%$B!!!!(B\\
%$B!!!!(B\\
%$B!!!!(B\\
%$B!!!!(B
%\vskip 5cm
}
\author{%
Yoshitaka {\sc Okumura}
\thanks{On leave of absence from Chubu University, Kasugai, 487-8501, Japan}
\thanks{
e-mail address: okum@bu.edu}\\
{\it Department of Physics, 
Boston University, Boston, MA 02215
}}
\date{}
\maketitle
\vskip 0.5cm
\begin{abstract}
{Noncommutative geometry(NCG) on the discrete space successfully reproduces
the Higgs mechanism of the spontaneously broken gauge theory, in which
the Higgs boson field is regarded as a kind of gauge field on the discrete
space. We could construct the generalized differential geometry(GDG) on the 
discrete space $M_4\times Z_N$ which is very close to NCG 
in case of $M_4\times Z_2$ and $M_4\times Z_3$. GDG is a direct 
generalization of the differential geometry on the ordinary
manifold into the discrete one. In this paper,
we attempt to construct the BRST invariant formulation
of spontaneously broken gauge 
theory based on GDG and obtain the BRST invariant Lagrangian with the
t'Hooft-Feynman gauge fixing term.}
\end{abstract}
%%%%%%%%%%%%%%%%%%%%%%%%%%%%%%%%%%%%%%%%%%%%%%%%%%%%%%%%%%%%%%%%%%%%%%%%%%%%%
\thispagestyle{empty}
%%%%%%%%%%%%%%%%%%%%%%%%%%%%%%%%%%%%%%%%%%%%%%%%%%%%%%%%%%%%%%%%%%%%%%%%%%%
%\setlength{\baselineskip}{0.85cm}
%\setlength{\baselineskip}{0.65cm}
\section{Introduction}
The standard model in particle physics has matched with all experimental data
conducted so far. Only ingredient which remains undetermined is the Higgs
boson field that causes the spontaneous breakdown of symmetry through
its vacuum expectation value. The next targets of future accelerators
such as LHC and Tevatron II are to detect the Higgs boson and supersymmetric
partners of particles and to find the breakthrough out of the standard model.
Thus, the great concerns about the Higgs boson field have been kept 
also from now.
\par 
Let us put emphasis on its similarity with the gauge boson as one of
the characteristic features of the Higgs boson in the standard model.
Regardless to say, the Higgs boson field is a boson field 
as the gauge field though it is scalar field. 
In addition, the Higgs field has the same type coupling
with fermions as the gauge fields and it has trilinear and quartic 
self-couplings as well as weak gauge fields. From these similarity between
the Higgs and gauge bosons, an idea that the Higgs boson may be a kind of
gauge boson comes out. In fact, as models to realize this idea, there has been
proposed the Kaluza-Klein model \cite{Man}, 
and Noncommutative geometry(NCG) \cite{Con}, \cite{MM} on the discrete space.
Especially, NCG approach does not require any extra physical mode and 
realizes the unified picture of gauge and Higgs fields as 
the generalized connection on the discrete space $M_4\times Z_2$.\par
Since the first formulation of NCG by Connes \cite{Con}, 
many versions of NCG  \cite{MM}-\cite{MO1} 
has appeared and succeeded to reconstruct the 
spontaneously broken gauge theories.
Morita and the present author \cite{MO1} proposed the generalized 
differential geometry (GDG) on the discrete space $M_4\times Z_2$
and reconstructed the Weinberg-Salam model. 
In this formulation on $M_4\times Z_2$ 
the extra differential one-form $\chi$ 
is introduced in addition to
the usual one-form $dx^\mu$ and so our formalism is the generalization
of the ordinary differential geometry on the compact manifold.
This formulation was generalized to GDG on the discrete space  
$M_4\times Z_{N}$ \cite{GUT}, \cite{LR} 
by introducing the extra one-forms $\chi_k(k=1,2\cdots N)$,
which generalization enabled us to reconstruct the left-right 
symmetric gauge theory\cite{LR}, 
SU(5) GUT\cite{GUT} and SO(10) GUT\cite{O10}
as spontaneously broken gauge theories 
on the discrete space $M_4\times Z_{N}$.
\par
It is also very important to reconstruct the gauge fixing  
and ghost terms in NCG in order to ensure the quantization of gauge theory.
Lee, Hwang and Ne'eman \cite{LHN} succeeded in incorporating
these terms in NCG in the matrix derivative approach 
based on the super-connection formalism\cite{SC}.
%proposed by Coquereaux et al in Ref.\cite{NCG}.
They obtained the BRS/anti-BRS transformation rules of the theory by
applying the horizontality condition\cite{HC} 
in the super-connection formalism
and constructed BRST invariant Lagrangian including the gauge fixing and
ghost terms. The present author subsequently applied their idea to 
the GDG formulation of gauge theory on $M_4\times Z_2$ \cite{BRSTO}
and obtained the BRST invariant formulation. 
We apply in this article the similar method to more general formulation
of GDG on $M_4\times Z_{N}$ \cite{FSTM} based on 
our formulation of GDG.
The last section is devoted to concluding remarks. 
%%%%%%%%%%%%%%%%%%%%%%%%%%%%%%%%%%%%%%%%%%%%%%%%%%%%%%%%%%%%%%%%%%%%%%%%%%%%%%
\section{ BRST transformation in generalized differential geometry}
The generalized differential geometry (GDG) on $M_4\times Z_N$ was formulated 
\cite{GUT}, \cite{LR}
to reconstruct the gauge invariant Lagrangian of 
the spontaneously broken gauge theories such as
the standard model, the left-right symmetric gauge theory and 
SU(5) and SO(10) grand unified theories.
In this section, we incorporate BRST transformation in GDG according to
the super field formalism of Bonora and Tonin\cite{BT}.
This formulation\cite{BRSTO} has already done 
in GDG on the discrete space 
$M_4\times Z_2$.
 Here, we generalize it to $M_4\times Z_N$ where 
$x_\mu$ and $n(n=1,2\cdots N)$ are arguments in $M_4$ and $Z_N$, respectively.
\par
Let us start with the equation of the generalized gauge field 
${\cal A}(x,n,\theta,\bar{\theta})$
written in one-form on the discrete space $M_4\times Z_N$:
\be
      {\cal A}(x,n,\theta,\bar{\theta})
      =\sum_{i}a^\dagger_{i}(x,n,\theta,\bar{\theta}){\boldsymbol d}
      a_i(x,n,\theta,\bar{\theta}) \label{2.1}
\ee
by adding the Grassmann numbers $\theta$
and $\bar{\theta}$ to $x_\mu$ and $n$ to produce the ghost and 
anti-ghost fields. 
The constituent $a_i(x,n,\theta,\bar{\theta})$ 
is the square-matrix-valued function
with a subscript $i$ which is a variable of the extra internal space. 
Now, we simply regard $a_i(x,n,\theta,\bar{\theta})$  as the more fundamental
field to construct gauge and Higgs fields because it has only
mathematical meaning and never appears in final stage.
The operator ${\boldsymbol{d}}$ in Eq.(\ref{2.1}) is the generalized exterior 
derivative defined as follows.
\bea 
   &&    {\boldsymbol d}=d + \sum_{k=1}^Nd_{\chi_k}+d\theta+ 
        d{\bar{\theta}}, \label{2.2a1}\\  
    && da_i(x,n,\theta,\bar{\theta}) =
   \partial_\mu a_i(x,n,\theta,\bar{\theta})dx^\mu, \label{2.2a2}\\
  && d_{\chi_k} a_i(x,n,\theta,\bar{\theta}) 
   % =\p_k a_i(x,n,\theta,\bar{\theta})\chi 
    =[-a_i(x,n,\theta,\bar{\theta})M_{nk} 
      + M_{nk}a_i(x,k,\theta,\bar{\theta})]\chi_k, \label{2.2a3}\\
  && d_\theta a_i(x,n,\theta,\bar{\theta})
       =\p_\theta a_i(x,n,\theta,\bar{\theta})d\theta  \label{2.2a4}\\
  && d_{\bar{\theta}} a_i(x,n,\theta,\bar{\theta})
    =\p_{\bar{\theta}}a_i(x,n,\theta,\bar{\theta})d\bar{\theta}, 
    \label{2.2a5}
\eea
where 
$dx^\mu$ is ordinary one-form basis, taken to be dimensionless, 
in Minkowski space 
$M_4$, and $\chi_k$ 
is the one-form basis, assumed to be also dimensionless, 
in the discrete space $Z_N$. $d_\theta,d_{\bar{\theta}}$ are also one-form
base in super-space.
We have introduced $x$-independent matrix $M_{nk}$ 
whose hermitian conjugation is given by $M_{nk}^\dagger=M_{kn}$. 
The matrix $M(y)$ turns out to determine the scale and pattern of 
the spontaneous breakdown of the gauge symmetry. 
Thus, the symmetry
breaking mechanism is encoded in the $d_\chi=\sum_{k=1}^Nd_{\chi_k}$ operation.
In order to find the explicit forms of gauge, Higgs fields and ghost fields
according to Eqs. (\ref{2.1}) and (\ref{2.2a1})$\sim$(\ref{2.2a5}), 
we need the following important algebraic rule of GDG.
\be
  \chi_kf(x,n,\theta,\bar{\theta})=f(x,k,\theta,\bar{\theta})\chi_k, 
  \label{2.4}
\ee
where $f(x,n,\theta,\bar{\theta})$ is a field defined 
on the discrete space such as
$a_i(x,n,\theta,\bar{\theta})$, gauge field, Higgs field , ghosts 
or fermion fields.
It should be noticed that Eq.(\ref{2.4}) never expresses 
the relation between
the matrix elements of $f(x,n,\theta,\bar{\theta})$ 
and $f(x,k,\theta,\bar{\theta})$ but insures the product between
the fields expressed in  differential form on the discrete space. 
Equation(\ref{2.4}) realizes the non-commutativity of our algebra 
in the geometry on the discrete space $M_4\times Z_N$. 
Inserting Eq.(\ref{2.2a1})$\sim$Eq.(\ref{2.2a5}) into Eq.(\ref{2.1})
and using Eq.(\ref{2.4}),
${\cal A}(x,n,\theta,\bar{\theta})$ is rewritten as
\be
 {\cal A}(x,n,\theta,\bar{\theta})={A}_\mu(x,n,\theta,\bar{\theta})dx^\mu
 +\sum_{k=1}^N{\mit\Phi}_{nk}(x,\theta,\bar{\theta})\chi_k
+{C}(x,n,\theta,\bar{\theta})d\theta
 +{{\bar C}}(x,n,\theta,\bar{\theta})d{\bar\theta}, \label{2.5}
\ee
where
\bea
&&    A_\mu(x,n,\theta,\bar{\theta}) = \sum_{i}a_{i}^\dagger(x,n,
  \theta,\bar{\theta})\partial_\mu a_{i}(x,n,\theta,\bar{\theta}), 
 \label{2.61}\\
&&     {\mit\Phi}_{nk}(x,\theta,\bar{\theta}) 
 = \sum_{i}a_{i}^\dagger(x,n,
  \theta,\bar{\theta})\,(-a_i(x,k,\theta,\bar{\theta})M_{nk} 
            + M_{nk}a_i(x,k,\theta,\bar{\theta})), \label{2.62}\\
&&  {C}(x,n,\theta,\bar{\theta})
  =\sum_{i}a_{i}^\dagger(x,n,\theta,\bar{\theta}) 
      \p_\theta a_{i}(x,n,\theta,\bar{\theta})
  \label{2.63}\\
&&  {\bar C}(x,n,\theta,\bar{\theta})
  =\sum_{i}a_{i}^\dagger(x,n,\theta,\bar{\theta}) 
      \p_{\bar\theta} a_{i}(x,n,\theta,\bar{\theta}).
  \label{2.64}\\
\eea
$A_\mu(x,n,\theta,\bar{\theta})$, ${\mit\Phi}_{nk}(x,\theta,\bar{\theta})$, 
$C(x,n,\theta,\bar{\theta})$ and 
${\bar C}(x,n,\theta,\bar{\theta})$ are identified with
the gauge field in the flavor symmetry, Higgs field,
ghost and anti-ghost fields, respectively. 
In order to identify  $A_\mu(x,n,\theta,\bar{\theta})$ 
with true gauge fields, the following conditions have to be imposed.
\bea
&&    \sum_{i}a_{i}^\dagger(x,n,\theta,\bar{\theta})
     a_{i}(x,n,\theta,\bar{\theta})= 1, 
  \label{2.7}
\eea
where $1$ in the right hand side is a unit matrix in the corresponding
internal space.
This equation is very important
often used in later calculations and may suggest that 
the variable $i$ might be an argument in the internal space 
because the definition of gauge field in Eq.(\ref{2.61}) is very 
similar to that in Berry phase \cite{Ber}. 
If we define the operator $\partial_k$ as
\be
\partial_ka_i(x,n)=-a_i(x,n)M_{nk}+M_{nk}a_i(x,k)
\ee
the Higgs field ${\mit\Phi}_{nk}(x,\theta,\bar{\theta})$
is written as
\be
{\mit\Phi}_{nk}(x,\theta,\bar{\theta})=\sum_ia_i^\dagger(x,n)
\partial_k a_i(x,n),
\ee
which is the same form as the ordinary gauge field 
$A_\mu(x,n,\theta,\bar{\theta})$ in Eq.(\ref{2.61}).
For later convenience, we define the following one-form
fields as
\bea
 &&  {\hat A}(x,n,\theta,\bar{\theta})=A_\mu(x,n,\theta,\bar{\theta})dx^\mu,
                     \label{2.6a1} \\
&& {\hat{\mit\Phi}}_{nk}(x,\theta,\bar{\theta})
={\mit\Phi}_{nk}(x,\theta,\bar{\theta})\chi_k,
                     \label{2.6a2} \\
&&      {\hat C}(x,n,\theta,\bar{\theta})
        ={C}(x,n,\theta,\bar{\theta})d\theta, \lab{2.6a3} 
&&      {\hat{\bar C}}(x,n,\theta,\bar{\theta}),
               ={\bar C}(x,n,\theta,\bar{\theta})d{\bar\theta}.\label{2.6a4} 
\eea
\par 
Before constructing the gauge covariant field strength, 
we address the gauge transformation \\
of $a_i(x,y,\theta,\bar{\theta})$  which is defined as 
\bea
&&      a^{g}_{i}(x,n,\theta,\bar{\theta})
        = a_{i}(x,n,\theta,\bar{\theta})g(x,n), 
\label{2.8}
\eea
where
$g(x,n)$ is the gauge function with respect to the corresponding
flavor unitary group.
Then, we can find from Eqs.(\ref{2.1}) and (\ref{2.8})
the gauge transformation of ${\cal A}(x,n,\theta,\bar{\theta})$ to be
\be
{\cal A}^g(x,n,\theta,\bar{\theta})=g^{-1}(x,n) 
{\cal A}(x,n,\theta,\bar{\theta})g(x,n)
 +g^{-1}(x,n){\boldsymbol d}g(x,n), \label{2.9}
\ee
where  as in Eq.(\ref{2.2a1})$\sim$Eq.(\ref{2.2a3}),
\bea 
   {\boldsymbol d}g(x,n)\a=(d+\sum_{k=1}^Nd_{\chi_{nk}}) g(x,n)
    =\p_\mu g(x,n)dx^\mu+\sum_{k=1}^N\p_{k}g(x,n)\chi_k 
          \label{2.10}
\eea
Using Eqs. (\ref{2.8})and (\ref{2.9}), 
we can find the gauge transformations of gauge, Higgs, ghost and anti-ghost
fields as
\bea
&&      A_\mu^g(x,n,\theta,\bar{\theta})=g^{-1}(x,n)
   A_\mu(x,n,\theta,\bar{\theta})g(x,n)+
                           g^{-1}(x,n)\p_\mu g(x,n),  
                           \label{2.11}\\
&&  {\mit\Phi}_{nk}^g(x,\theta,\bar{\theta})
=g^{-1}(x,n){\mit\Phi}_{nk}(x,\theta,\bar{\theta})
g(x,k)+  g^{-1}(x,n)\p_{k}g(x,n), \label{2.12}\\
&&      C^g(x,n,\theta,\bar{\theta})=g^{-1}(x,y)C(x,n,\theta,\bar{\theta})
   g(x,n),  \label{2.13}\\
&&      {\bar C}^g(x,n,\theta,\bar{\theta})=g^{-1}(x,n)
      {\bar C}(x,n,\theta,\bar{\theta})g(x,n),  \label{2.13a}
\eea
Equation(\ref{2.12}) is very similar to Eq.(\ref{2.11}) that is 
the gauge transformation of
the genuine gauge field $A_\mu(x,n,\theta,\bar{\theta})$ and so 
it strongly indicates that the Higgs field is a kind of gauge field
on the discrete space $M_4\times Z_N$. From Eq.(\ref{2.10}),  
Eq.(\ref{2.12})  is rewritten as
\be
       {\mit\Phi}_{nk}^g(x,\theta,\bar{\theta})+M_{nk}=g^{-1}(x,n)
       ({\mit\Phi}_{nk}(x,\theta,\bar{\theta})+M_{nk})g(x,k),
                              \label{2.14}\\
\ee
which makes obvious that 
\be
H_{nk}(x,\theta,\bar{\theta})={\mit\Phi}_{nk}(x,\theta,\bar{\theta})
+M_{nk} \label{2.15}
\ee
is un-shifted Higgs field whereas ${\mit\Phi}_{nk}(x,\theta,\bar{\theta})$ 
denotes shifted one with the vanishing vacuum expectation value.
Equations (\ref{2.13}) and (\ref{2.13a}) show that ghost and anti-ghost
fields are transformed as the adjoint representation.
\par
In addition to the algebraic rules in Eq.(\ref{2.2a1})$\sim$(\ref{2.2a5}) 
we add one more important rule that
\be
         d_{\chi_l}(M_{nk}\chi_k)=(M_{nl}\chi_l)\wedge (M_{nk}\chi_k)=
         M_{nl}M_{lk}\chi_l\wedge\chi_k,
         \label{2.16}
\ee
and in addition,  whenever the $d_{\chi_k}$ operation jumps over 
$M_{nl}\chi_l$, a minus sign is attached.
For example,
\be
    d_{\chi_k}\left(M_{nl}\chi_la(x,n)\right)=
    \left(d_{\chi_k}M_{nl}\chi_l\right)a(x,n)
    - M_{nl}\chi_l\wedge \left(d_{\chi_k}a(x,n)\right).
\label{2.17b}
\ee
which together with Eq.(\ref{2.2a3}) yields  the nilpotency
of $d_\chi=\sum_{k=1}^Nd_{\chi_k}$ and then the nilpotency of 
the generalized exterior derivative 
$\boldsymbol d$ 
under the natural conditions that 
\bea
\a dx^\mu\wedge\chi_k=-\chi_k\wedge dx^\mu, \hskip 1cm
   dx^\mu\wedge d\theta=-d\theta\wedge dx^\mu, \hskip 1cm
   dx^\mu\wedge d{\bar\theta}=-d{\bar\theta}\wedge dx^\mu, \nonumber\\
\a \chi_k\wedge d\theta=-d\theta\wedge \chi_k \hskip 1cm  
   \chi_k\wedge d{\bar\theta}=-d{\bar\theta}\wedge \chi_k \nonumber\\
\a d{\theta}\wedge d{\bar\theta}=d{\bar\theta}\wedge d\theta, \hskip 1cm
  \p_\theta\p_{\bar\theta}=-\p_{\bar\theta}\p_\theta.
      \label{2.16a}
\eea
It should be noted that $\chi_k\wedge \chi_l$ is independent of
$\chi_l\wedge \chi_k$ so that $\chi_k\wedge \chi_l\ne \chi_l\wedge\chi_k$. 
This independence is due to the noncommutative property of our generalized
differential geometry.
For the proof of nilpotency of $d_\chi$, see \cite{LR}.
With these considerations we can construct the gauge covariant field
strength:
\be
  {\cal F}(x,n,\theta,\bar{\theta})
  = {\boldsymbol d}{\cal A}(x,n,\theta,\bar{\theta})
  +{\cal A}(x,n,\theta,\bar{\theta})\wedge{\cal A}(x,n,\theta,\bar{\theta})
\label{2.18}
\ee
From Eqs.(\ref{2.9}) and (\ref{2.10}) we can easily find the gauge 
transformation of ${\cal F}(x,n,\theta,\bar{\theta})$ as
\be
         {\cal F}^g(x,n,\theta,\bar{\theta})
         =g^{-1}(x,n){\cal F}(x,n,\theta,\bar{\theta})g(x,n).  \label{2.20}
\ee
\par
Here, according to Bonora and Tonin\cite{BT} we impose the horizontality
condition \cite{HC} on ${\cal F}(x,n,\theta,\bar{\theta})$ that
\be
    {\cal F}(x,n,\theta,\bar{\theta})|_{\theta={\bar\theta}=0}
    =F(x,n), \label{2.20a}
\ee
where $F(x,n)$ is the generalized field strength not accompanying
one-form base $d\theta$ and $d{\bar\theta}$.
Equation(\ref{2.20a}) yields the conditions that
\bea
&&  d_\theta {\hat {A}}(x,n)+d{\hat {C}}(x,n)+{\hat {A}}(x,n)\wedge 
       {\hat {C}}(x,n)
       +{\hat {C}}(x,n)\wedge {\hat {A}}(x,n)=0, \label{2.20a1}\\
&&  d_{\bar\theta} {\hat A}(x,n)+d{\hat{\bar C}}(x,n)
   +{\hat A}(x,n)\wedge {\hat{\bar C}}(x,n)
       +{\hat{\bar C}}(x,n)\wedge {\hat A}(x,n)=0, \label{2.20a2}\\
&&  d_\theta{\hat {{\mit\Phi}}}_{nk}(x)+d_{\chi_k}{\hat {C}}(x,n)
       +{\hat {\mit\Phi}_{nk}}(x)\wedge {\hat {C}}(x,n)
       +{\hat {C}}(x,n)\wedge {\hat {\mit\Phi}_{nk}}(x)=0, \label{2.20a3}\\
&&  d_{\bar{\theta}} {\hat {\mit\Phi}_{nk}}(x)+d_\chi{\hat{\bar {C}}}(x,n)
   +{\hat {\mit\Phi}_{nk}}(x)\wedge {\hat{ \bar{C}}}(x,n)
       +{\hat{\bar C}}(x,n)\wedge {\hat {\mit\Phi}_{nk}}(x)=0, \label{2.20a4}\\
&& d_\theta {\hat C}(x,n)+{\hat C}(x,n)\wedge{\hat C}(x,n)=0,
                                       \label{2.20a5},\\
&& d_{\bar\theta} {\hat{\bar C}}(x,n)
    +{\hat{\bar C}}(x,n)\wedge{\hat{\bar C}}(x,n)=0,  \label{2.20a6},\\
&& d_{\bar\theta} {{\hat C}}(x,n)+d_{\theta} {\hat{\bar C}}(x,n)
      +{{\hat C}}(x,n)\wedge{\hat{\bar C}}(x,n)
      +{\hat{\bar C}}(x,n)\wedge{\hat C}(x,n)=0, \label{2.20a7}
\eea
which determine the BRST/anti-BRST transformations of each field
together with the definitions that
\be
    d_\theta{\hat{\bar C}}(x,n)=-{\hat B}(x,n),\hskip 2cm 
    d_{\bar\theta}{\hat C}(x,n)=-{\hat{\bar B}}(x,n). \label{2.20a8}
\ee
It should be noticed that nilpotencies of $d_\theta$ and $d_{\bar\theta}$ 
are consistent with Eqs.(\ref{2.20a1})$\sim$(\ref{2.20a8})
and for example, $d_\theta{\hat{ C}}(x,n)=\partial_\theta d\theta
{\hat{C}}(x,n)=-\partial_\theta {{C}}(x,n)d\theta\wedge d\theta$
because $\theta$ and $C(x,n)$ are Grassmann numbers. Thus,
\be
    \partial_\theta{{\bar C}}(x,n)={ B}(x,n),\hskip 2cm 
    \partial_{\bar\theta}{ C}(x,n)={{\bar B}}(x,n). \label{2.20a9}
\ee
\par
From Eqs.(\ref{2.20a1}) and (\ref{2.20a2}), the BRST/anti-BRST transformations
of $A_\mu(x,n)$ follows as
\bea
    &&     \partial_\theta A_\mu(x,n)={\cal D}_\mu C(x,n)=
           \partial_\mu C(x,n)+[A(x,n), C(x,n)],  \label{2.20c1}\\
    &&     \partial_{\bar\theta} A_\mu(x,n)={\cal D}_\mu {\bar C}(x,n)=
           \partial_\mu {\bar C}(x,n)+[A(x,n), {\bar C}(x,n)]. \label{2.20c2}
\eea
From Eq.(\ref{2.6a2})$\sim$Eq.(\ref{2.6a4}), the BRST/anti-BRST 
transformation of the Higgs field are rewritten as
\bea
&&   \partial_\theta {\mit\Phi}_{nk}(x)=
   \partial_{k} C(x,n)+{\mit\Phi}_{nk}(x)C(x,k)-C(x,n){\mit\Phi}_{nk}(x),
                             \label{{2.20b1}}\\
&&   \partial_{\bar\theta} {\mit\Phi}_{nk}(x)=\p_{k} {\bar C}(x,n)
    +{\mit\Phi}_{nk}(x){\bar C}(x,k)-{\bar C}(x,n){\mit\Phi}_{nk}(x),
                             \label{{2.20b2}}
\eea
which by use of $H_{nk}(x)={\mit\Phi}_{nk}(x)+M_{nk}$ lead to
\bea
&&   \partial_\theta H_{nk}(x)=H_{nk}(x)C(x,k)-C(x,n)H_{nk}(x),
                             \label{2.20b3}\\
&&   \partial_{\bar\theta} H_{nk}(x)
           =H_{nk}(x){\bar C}(x,k)-{\bar C}(x,n)H_{nk}(x).
                             \label{2.20b4}
\eea
Equations (\ref{2.20b3}) and (\ref{2.20b4}) are the usual BRST/anti-BRST
transformation of the Higgs field. 
From Eqs.(\ref{2.20a5}) and (\ref{2.20a6}), the BRST and anti-BRST 
transformations of ghost and anti-ghost fields are obtained, respectively.
\bea
     &&   \partial_\theta C(x,n)=-C(x,n)C(x,n), \label{2.20c3}\\
     &&   \partial_{\bar\theta} {\bar C}(x,n)
      =-{\bar C}(x,n){\bar C}(x,n), \label{2.20c4}
\eea
and from Eq.(\ref{2.20a7}), the restriction about Nakanishi-Lautrup field
follows as
\be
     B(x,n)+{\bar B}(x,n)=-C(x,n){\bar C}(x,n)-{\bar C}(x,n)C(x,n).
     \label{2.20c5}
\ee
With these equations, we can proceed to obtain the BRST invariant
Lagrangian of gauge theory.
\par
BRST invariant \ymh is obtained by
\bea
     {\cal L}_{\hbox{\rm\tiny YMH}}&=&-\sum_{n=1}^N\frac1{g_n^2}
     {\rm Tr}<F(x,n), F(x,n)>\nonumber\\
             &&+\sum_{n=1}^N\frac1{g_n^2}i\partial_\theta\partial_{\bar\theta}
  {\rm Tr}<{\cal A}(x,n,\theta,{\bar\theta}), 
  {\cal A}(x,n,\theta,{\bar\theta})>|_{\theta={\bar\theta}=0} \nonumber\\
 &&+\sum_{n=1}^N\frac1{g_n^2}\frac\alpha2
           {\rm Tr}<{\hat B}(x,n,\theta,{\bar\theta}), 
   {\hat B}(x,n,\theta,{\bar\theta})>|_{\theta={\bar\theta}=0},
                \label{2.19}
\eea
where $g_n$ is a constant relating 
to the coupling constant of the flavor gauge field and
Tr denotes the trace over internal symmetry matrices. 
In order to express \ymh let us denote the explicit expressions of
the field strength $F(x,n)$.
The algebraic rules defined in Eqs.(\ref{2.2a1})$\sim$(\ref{2.2a3}), 
(\ref{2.4}) and (\ref{2.7}) yield
\bea
F(x,n) &=& { 1 \over 2}F_{\mu\nu}(x,n)dx^\mu \wedge dx^\nu + 
    \sum_{k\ne n}D_\mu{\mit\Phi}_{nk}(x)dx^\mu \wedge \chi_k \label{2.25}\\
    && + \sum_{k\ne n}V_{nk}(x)\chi_k \wedge \chi_n 
                 \label{2.21}
\eea
where
\bea
&&  F_{\mu\nu}(x,n)=\p_\mu A_\nu (x,n) - \p_\nu A_\mu (x,n) 
               +[A_\mu(x,n), A_\mu(x,n)], \label{2.221}\\
&&  D_\mu {\mit\Phi}_{nk}(x)=\p_\mu {\mit\Phi}_{nk}(x)  
+ A_\mu(x,n)(M_{nk} + {\mit\Phi}_{nk}(x))\nonumber\\
&&     \hskip 6.5cm           
 -({\mit\Phi}_{nk}(x)+M_{nk})A_\mu(x,k),\label{2.222} \\  
&& V_{nk}(x)= ({\mit\Phi}_{nk}(x) + M_{nk})({\mit\Phi}_{kn}(x) + M_{kn}) -
             Y_{nk}(x) \hskip 0.5cm{\rm for} \quad k \ne n,
\label{2.223}
\eea
$Y_{nk}(x)$ in Eq.(\ref{2.223}) is auxiliary field and expressed as 
\be
  Y_{nk}(x)= \sum_{i}a_{i}^\dagger(x,n)M_{nk}M_{kn}a_{i}(x,n),
 \label{2.23}
\ee
which may be independent or dependent of ${\mit\Phi}_{nk}(x)$ 
and/or may be a constant field.
\par
In order to get the explicit expression of $L_{\rm YMH}$ 
in Eq.(\ref{2.19})
we have to determine the metric structure of one-forms.
\bea
\a <dx^\mu, dx^\nu>=g^{\mu\nu},\quad 
g^{\mu\nu}={\rm diag}(1,-1,-1,-1),\nonumber\\
\a <\chi_k, \chi_l>=-\delta_{kl}, \hskip 1cm 
<d\theta, d\theta>=<d{\bar\theta}, d{\bar\theta}>=1,
\label{2.25a} 
\eea
with the vanishing values of other combinations.
From Eqs.(\ref{2.21})$\sim$(\ref{2.223}), the first term of
Eq.(\ref{2.19}) that is denoted by 
${\cal L}_{\hbox{\rm\tiny YMH}}$ is written as
\bea
{\cal L}_{\hbox{\rm\tiny YMH}}
&=&-{\rm Tr}\sum_{n=1}^N{1\over 2g^2_n}
F_{\mu\nu}^{\dag}(x,n)F^{\mu\nu}(x,n)\nonumber\\
&&+{\rm Tr}\sum_{n,k=1}^N{1\over g_{n}^2}
    (D_\mu {\mit\Phi}_{nk}(x))^{\dag}D^\mu {\mit\Phi}_{nk}(x)  \nonumber\\
&& -{\rm Tr}\sum_{n,k=1}^N{1\over g_{n}^2}
        V^{\dag}_{nk}(x)V_{nk}(x),
\label{2.26}
\eea
where the third term in the right hand side of Eq.(\ref{2.26}) 
is the potential term of Higgs particle.
The second and third terms of Eq.(\ref{2.19}) give the ghost term 
${\cal L}_{\rm GH}$ and the gauge fixing term ${\cal L}_{\rm GF}$,
respectively. 
The calculation of the second term in Eq.(\ref{2.19}) proceeds as
\bea
&&\partial_\theta\partial_{\bar\theta}
  {\rm Tr}<{\cal A}(x,n,\theta,{\bar\theta}), 
  {\cal A}(x,n,\theta,{\bar\theta})>|_{\theta={\bar\theta}=0}=\nonumber\\
  &&\hskip1cm-\partial_\theta\partial_{\bar\theta}
  {\rm Tr}\left[A^\mu(x,n,\theta,{\bar \theta})A_\mu(x,n,\theta,{\bar \theta})
    +\sum_{k=1}^N{\mit\Phi}_{kn}(x,\theta,{\bar\theta})
   {\mit\Phi}_{nk}(x,\theta,{\bar\theta})\right.\nonumber\\
&& \hskip2.5cm  \left.+C(x,n,\theta,{\bar\theta})C(x,n,\theta,{\bar\theta})
   +{\bar C}(x,n,\theta,{\bar\theta}){\bar C}(x,n,\theta,{\bar\theta})
   {\vphantom{\frac12}}\right]_{\theta={\bar\theta}=0},
\eea
because
$A^\mu(x,n,\theta,{\bar \theta})$, $C(x,n,\theta,{\bar \theta})$ and
${\bar C}(x,n,\theta,{\bar \theta})$ are anti-Hermite and 
$\Phi_{nk}(x,\theta,{\bar\theta})^\dagger=\Phi_{kn}(x,\theta,{\bar\theta})$
together with
the internal product $<\chi_k,\chi_l>=-\delta_{kl}$.
It should be noted that
\bea
&&\partial_\theta \left\{C(x,n,\theta,{\bar \theta})
C(x,n,\theta,{\bar \theta})\right\}=0,\\
&&\partial_{\bar\theta}\left\{ {\bar C}(x,n,\theta,{\bar \theta})
  {\bar C}(x,n,\theta,{\bar \theta})\right\}=0,
\eea
due to the nilpotency of BRST/anti-BRST transformation.
Then, according to the BRST/anti-BRST transformations 
in Eqs.(\ref{2.20c1})$\sim$(\ref{2.20c5}), 
${\cal L}_{\rm GH}$ is expressed as
\bea
{\cal L}_{\hbox{\rm\tiny GH}}=\a 2i\sum_{n=1}^N\frac1{g_n^2}{\rm Tr}
\p_\mu{\bar C}(x,n){\cal D}^\mu C(x,n)\nonumber\\
\a +i\sum_{n=1}^N\frac1{g_n^2}\sum_{k\ne n}^N
 {\rm Tr}(\p_n {\bar C}(x,k){\cal D}_kC(x,n)+
    \p_k {\bar C}(x,n){\cal D}_nC(x,k) ),
            \label{2.27}
\eea
where 
\bea
    {\cal D}^\mu C(x,n)&=&\p^\mu C(x,n)+[A^\mu(x,n), C(x,n)], \label{2.271}\\
  \p_k{\bar C}(x,n)&=&-{\bar C}(x,n)M_{nk}+M_{nk}{\bar C}(x,k),
                                      \label{2.270}\\
    {\cal D}_k C(x,n)&=&\partial_k C(x,n)
    +{\mit\Phi}_{nk}(x)C(x,k)-C(x,n){\mit\Phi}_{nk}(x)\nonumber\\
                     &=&H_{nk}(x)C(x,k)-C(x,n)H_{nk}(x)
               \label{2.272} 
\eea
and ${\cal L}_{\hbox{\rm\tiny GF}}$ as
\bea
   {\cal L}_{\hbox{\rm\tiny GF}}
   =\a\frac\alpha2\sum_{n=1}^N\frac1{g_n^2}{\rm Tr}B(x,n)^2
   -2i \sum_{n=1}^N\frac1{g_n^2}{\rm Tr}
               \p_\mu B(x,n)A^\mu(x,n)\nonumber\\
             -\a i\sum_{n=1}^N\frac1{g_n^2}
            \sum_{k\ne n}^N{\rm Tr}
             \left(\p_n B(x,k){\mit\Phi}_{nk}(x)
             +{\mit\Phi}_{kn}(x)\p_k B(x,n)\right).
               \label{2.28}
\eea
If we note the Hermitian conjugate conditions that 
\bea
   \a \left(\p_k {\bar C}(x,n)\right)^\dagger=\p_n {\bar C}(x,k), \hskip 1cm
   \left({\cal D}_k {C}(x,n)\right)^\dagger={\cal D}_n {C}(x,k) \nonumber\\
   \a \left(\p_k {B}(x,n)\right)^\dagger=-\p_n {B}(x,k)
                \label{2.281}
\eea
because of $B(x,n)^\dagger=B(x,n)$, $C(x,n)^\dagger=-C(x,n)$ and
${\bar C}(x,n)^\dagger=-{\bar C}(x,n)$, we easily find the Hermiticity of
Eqs.(\ref{2.27}) and (\ref{2.28}).\par
In next two sections, two special models, SU(2) Higgs-Kibble gauge model
and the standard model are reconstructed according to the general 
framework in this section.
\par
%%%%%%%%%%%%%%%%%%%%%%%%%%%%%%%%%%%%%%%%%%%%%%%%%%%%%%%%%%%%%%%%%%%%%%%%%%%
\section{ Application to SU(2) Higgs-Kibble gauge model}
In this section we apply the previous results to the spontaneously broken
SU(2) gauge model. 
We need the discrete space $M_4\times Z_2(N=2)$ to reproduce the 
Higgs mechanism in the SU(2) Higgs Kibble model.
Let us first assign  the fields on discrete space $M_4\times Z_2$ 
to the physical fields. For gauge fields,
\bea
\a A_\mu(x,1)=-{i \over 2}\sum_{i=1}^3\tau^i A^i_\mu(x),\nonumber\\
\a A_\mu(x,-)=0,\label{3.1}
\eea
where $A_\mu^i(x)$  denotes SU(2) adjoint
gauge fields. $\tau^i(i=1,2,3)$ are Pauli matrices 
The Higgs field is
assigned as
\bea
\a {\mit\Phi}_{12}(x)={\mit\Phi}_{21}(x)^\dagger=
\left(
\matrix{
         {\phi^{0}}^\ast & \phi^+ \cr
         -\phi^-   & \phi^0 \cr
}
\right),
\nonumber\\
\a M_{12}=M_{21}^\dagger=
\left(
\matrix{
  \mu & 0 \cr
  0   & \mu \cr
}\right),\label{3.2}
\eea
where $M_{12}$ must be chosen to give the correct symmetry breakdown.
For ghost and anti-ghost fields which correspond with gauge fields 
in Eq.(\ref{3.1}) we take 
\bea
\a C(x,1)=-{i \over 2}\sum_{i=1}^3\tau^i C^i(x)
           ,\nonumber\\
\a C(x,2)=0\label{3.3}
\eea
and
\bea
\a {\bar C}(x,1)=-{i \over 2}\sum_{i=1}^3\tau^i {\bar C}^i(x)
            ,\nonumber\\
\a {\bar C}(x,2)=0.\label{3.4}
\eea
Also for the Nakanishi-Lautrup field, we assign
\bea
\a B(x,1)={1 \over 2}\sum_{i=1}^3\tau^i B^i(x)
           ,\nonumber\\
\a B(x,2)=0\label{3.5}
\eea
and
\bea
\a {\bar B}(x,1)={1 \over 2}\sum_{i=1}^3\tau^i {\bar B}^i(x)
            ,\nonumber\\
\a {\bar B}(x,2)=0\label{3.6}
\eea
because $\p_{ \theta} {\bar C}^i=i{B}^i$ and 
$\p_{ \theta} {\bar C}^0=i{B}^0$.
We can take the gauge transformation functions as
\bea
\a g(x,1)=g(x),\;  g(x)\in {\rm SU(2)},\nonumber\\
\a g(x,2)=1. \label{3.7}
\eea
The auxiliary field $Y(x,n)(n=1,2)$ becomes unit matrix because of 
the assignments of $M_{nk}$ in Eq.(\ref{3.2}). 
\begin{eqnarray}
&& Y(x,1)=\sum_ia^\dagger_i(x,1)M_{12}M_{21}a_i(x,1)=
       \sum_ia^\dagger_i(x,1)a_i(x,1)=1^{24},\nonumber\\
&& Y(x,2)=\sum_ia^\dagger_i(x,2)M_{21}M_{12}a_i(x,2)=
       \sum_ia^\dagger_i(x,2)a_i(x,2)=1^{24}, \label{3.7B}
\end{eqnarray}
on account of Eq.(\ref{2.7})\par
With these considerations, we can obtain $L_{\hbox{\rm\tiny YMH}}$, 
$L_{\hbox{\rm \tiny GH}}$ and $L_{\hbox{\rm\tiny GF}}$
in Eqs.(\ref{2.26}), (\ref{2.27}) and (\ref{2.28}), respectively.
After the definition of the Higgs doublet 
$\ds{h=\left(\matrix{\phi^+\cr\phi^0+\mu}\right)}$ and
the following rescaling of fields
\be
 A_\mu^i(x)  \rightarrow g A_\mu^i(x),\hskip 1.5cm
 h(x) \rightarrow{g_{\hbox{\rm \tiny H}}}h(x), \label{3.8}
\ee
with $g=g_1$ and $g_{\hbox{\rm \tiny H}}=\sqrt{g_1^2g_2^2/2(g_1^2+g_2^2)}$,
we find  the standard \ymh for the Higgs Kibble gauge model.
\bea
{\cal L}_{\hbox{\rm\tiny YMH}}&=&
  -{1\over 4}\sum_{i=1}^3 F_{\mu\nu}^i(x)\cdot F^{i\mu\nu}(x)\nonumber\\
  &&+(D_\mu h(x))^{\dag}(D^\mu h(x)) 
-\lambda(\,h^{\dag}(x)\,h(x)-\mu^2)^2, \label{3.9}
\eea
where
\bea
&& F_{\mu\nu}^i (x) = \partial_\mu A_\nu^i(x)-\partial_\nu A_\mu^i(x)
       +g\,\epsilon^{ijk}A_\mu^j(x) A_\nu^k(x), \label{3.10}\\
&& D_\mu h(x)=\left[\partial_\mu-{i \over 2}g\,\sum_{i=1}^3 
             \tau^i\cdot  A^i_\mu(x)
\right]h(x),\label{3.11}
\eea
with $\lambda=g_{\hbox{\rm \tiny H}}^2$ and the rescaling of 
$\mu\rightarrow \sqrt{g_{\hbox{\rm \tiny H}}}\mu$.
Equation(\ref{3.9}) expresses \ymh of the Higgs-Kibble gauge theory 
with the symmetry SU(2) spontaneously broken to global SU(2).
\par
Let us move to the ghost and gauge fixing terms expressed in Eqs.(\ref{2.27}) 
and (\ref{2.28}). 
For simplicity, hereafter we abbreviate the argument $x$ 
in the respective fields.
After the same rescaling of ghost and Nakanishi-Lautrup fields as
in Eq.(\ref{3.8})
we get the gauge fixing term ${\cal L}_{\hbox{\rm\tiny GF}}$ 
in Eq.(\ref{2.27}) as
\be
  {\cal L}_{\hbox{\rm\tiny GF}}= \sum_{i=1}^3\left\{
        \frac\alpha2{B_i}^2+
        B_i\left(\partial^\mu A_\mu^i
        +m_{\hbox{\rm\tiny W}}\phi^i\right)\right\}
         , \label{3.13}
\ee
where $m_{\hbox{\rm\tiny{W}}}$ is the gauge boson mass and
$\phi^i(i=1,2,3)$ is given by the following parametrization of $h$.
\be
   h=\frac1{\sqrt{2}}\left(\psi+v+i\sum_{i=1}^3\tau^i\phi^i\right)
   \left(\matrix{0\cr
                 1\cr}\right)
 \label{3.16}
\ee
with $v=\sqrt{2}\mu$.
The equations of motion eliminate 
the Nakanishi-Lautrup fields from Eq.(\ref{3.13}), which yields
\be
  {\cal L}_{\hbox{\rm\tiny{GF}}}=
        -\sum_{i=1}^3\frac1{2\alpha}\left(\partial^\mu A_\mu^i+
        m_{\hbox{\rm\tiny{W}}}\phi^i\right)^2
         . \label{3.17}
\ee
Eq.(\ref{3.17}) enables us to obtain the gauge fixed Lagrangian
with the 't Hooft-Feynman gauge\cite{tH} when $\alpha=1$.
\par
With the same notations as in Eq.(\ref{3.17}) 
we get the explicit expression of ghost terms in Eq.(\ref{2.27}) as follows:
\bea
       {\cal L}_{\hbox{\rm\tiny{GH}}}&=& 
           -i\sum_{i=1}^3\left(\partial^\mu{\bar C}^i{\cal D}_\mu C^i
                 -m_{\hbox{\rm\tiny{W}}}^2{\bar C}^iC^i\right) \nonumber\\
          &&+i\frac{gm_{\hbox{\rm\tiny{W}}}}{2}\sum_{i=1}^3
          \left({\bar C}^iC^i\psi+f^{ijk}{\bar C}^i\phi^jC^k\right),
           \label{3.18}
\eea
where ${\cal D}_\mu C^i=\partial_\mu C^i
+g\epsilon^{ijk}A_\mu^jC^k$.
The ghost fields become massive and
the new interaction terms between ghosts and
Higgs fields appear. This is natural because the Higgs field is
a member of the generalized connection in GDG on the discrete space
in the same way as the gauge field $A_\mu$.\par
It should be noted that our definitions of the gauge fixing and ghost terms
are equal to 
\be
{\cal L}_{\hbox{\rm\tiny GH}+\hbox{\rm\tiny GF}}
=-i\partial_\theta\left[\sum_{i=1}^3{\bar C}^i\left(
\partial^\mu A^i_\mu+m_{\hbox{\rm\tiny W}}\phi^i+\frac12\alpha B^i    
\right)\right]
\label{5.1}
\ee
for the SU(2) Higgs-Kibble model. 
This type of prescription to determine the gauge fixing condition was proposed
by Kugo and Uehara \cite{KU}.
%%%%%%%%%%%%%%%%%%%%%%%%%%%%%%%%%%%%%%%%%%%%%%%%%%%%%%%%%%%%%%%%%%%%%%%%%%%%
\section{ Application to the standard model}
The reconstruction of the standard model in GDG was completely performed in 
\cite{FSTM} by adopting the discrete space $M_4\times Z_2$ on which
the fermion fields are represented as vectors in 24 dimensional internal 
space including weak isospin, hypercharge, color and generation indices.
Corresponding to this fermion representation,  gauge fields, Higgs boson,
ghost fields, Nakanishi-Lautrup fields are expressed in $24\times 24$
matrix forms as generators in 24-dimensional space. Here, we omit the color
gauge field because it does not bring any significant difference from our
results.
 \par
\bea
&& A_\mu(x,1)=-{i \over 2}\left(\sum_{i=1}^3\tau^i A^i_\mu(x)\otimes1^4
      +a\;B_\mu(x)\right)\otimes1^3,\nonumber\\
&& A_\mu(x,2)=-{\frac i2}\,b\;B_\mu(x)\otimes1^3,\label{4.1}
\eea
where $A_\mu^i(x)$  denotes SU(2) adjoint
gauge field and $B_\mu(x)$ is U(1) gauge field. 
$a$ and $b$ are the U(1) hypercharge matrices corresponding to 
the left and right handed fermions expressed as vectors 
in 24-dimensional space, respectively and are denoted as
\bea
&& a={\rm diag}\;\left(\frac13,\frac13,\frac13, -1, 
            \frac13,\frac13,\frac13,-1\right)\nonumber\\
&&  b={\rm diag}\;\left(\frac43,\frac43,\frac43,0,
            -\frac23,-\frac23,-\frac23,-2\right).
            \label{4.1a}
\eea
The Higgs field is
assigned in the same way as in Eq.(\ref{3.2}) by making it 
a $24\times 24$ matrix. The symmetry breaking
function $M_{nk}$ is also given in the same way.
\bea
\a {\mit\Phi}_{12}(x)={\mit\Phi}_{21}(x)^\dagger=
\left(
\matrix{
         {\phi^{0}}^\ast & \phi^+ \cr
         -\phi^-   & \phi^0 \cr
}
\right)\otimes1^{12},
\nonumber\\
\a M_{12}=M_{21}^\dagger=
\left(
\matrix{
  \mu & 0 \cr
  0   & \mu \cr
}\right)\otimes1^{12}.\label{4.2}
\eea
For ghost and anti-ghost fields which correspond with gauge fields 
in Eq.(\ref{4.1}) we take 
\bea
&& C(x,1)=-{i \over 2}\left(\sum_{i=1}^3\tau^i C^i(x)\otimes1^4
      +a\;C^0(x)\right)\otimes1^3,\nonumber\\
&& C(x,2)=-{\frac i2}\,b\;C^0(x)\otimes1^3,\label{4.3}
\eea
and
\bea
&& {\bar C}(x,1)=-{i \over 2}\left(\sum_{i=1}^3\tau^i {\bar C}^i(x)\otimes1^4
      +a\;{\bar C}^0(x)\right)\otimes1^3,\nonumber\\
&& {\bar C}(x,2)=-{\frac i2}\,b\;{\bar C}^0(x)\otimes1^3,\label{4.4}
\eea
Also for the Nakanishi-Lautrup field, we assign
\bea
&& B(x,1)={1 \over 2}\left(\sum_{i=1}^3\tau^i B^i(x)\otimes1^4
      +a\;B^0(x)\right)\otimes1^3,\nonumber\\
&& B(x,2)={\frac 12}\,b\;B^0(x)\otimes1^3,\label{4.5}
\eea
and
\bea
&& {\bar B}(x,1)={1 \over 2}\left(\sum_{i=1}^3\tau^i {\bar B}^i(x)\otimes1^4
      +a\;{\bar B}^0(x)\right)\otimes1^3,\nonumber\\
&& {\bar B}(x,2)={\frac 12}\,b\;{\bar B}^0(x)\otimes1^3,\label{4.6}
\eea
because $\p_{ \theta} {\bar C}^i=i{B}^i$ and 
$\p_{ \theta} {\bar C}^0=i{B}^0$.
We can take the gauge transformation functions as
\bea
\a g(x,1)=g(x)e^{ia\alpha(x)},\;  g(x)\in {\rm SU(2)},
       \;e^{ia\alpha(x)}\in{\rm U(1)},\nonumber\\
\a g(x,2)=e^{ib\alpha(x)},\;e^{ib\alpha(x)}\in{\rm U(1)}. \label{4.7}
\eea
The auxiliary field $Y(x,n)(n=1,2)$ becomes unit matrix because of 
the assignments of $M_{nk}$ in Eq.(\ref{3.2}). 
With these considerations, we can obtain $L_{\hbox{\rm\tiny YMH}}$, 
$L_{\hbox{\rm \tiny GH}}$ and $L_{\hbox{\rm\tiny GF}}$
in Eqs.(\ref{2.26}), (\ref{2.27}) and (\ref{2.28}), respectively.
\par
After the rescaling of fields
\bea
&&A_\mu^i(x)  \rightarrow g A_\mu^i(x),\hskip 1.0cm
B_\mu(x)    \rightarrow g'B_\mu(x), \label{4.8}\\
&&h(x) \rightarrow{g_{\hbox{\rm \tiny H}}}h(x), 
\eea
with 
\bea
&&g^2=\frac {g_1^2}{12},\nonumber\\ 
&& {g'}^2=\frac{2g_1^2g_2^2}{3g_2^2{\rm Tr} a^2+
      3g_1^2{\rm Tr}b^2}=\frac{g_1^2g_2^2}{16g_1^2+4g_2^2},\nonumber\\
 &&g_{\hbox{\rm \tiny H}}^2=\frac{g_1^2g_2^2}{24(g_1^2+g_2^2)}.
\eea
we find  the \ymh for the Standard model.
\bea
{\cal L}_{\hbox{\rm\tiny YMH}}&=&
  -{1\over 4}\sum_{i=1}^3 F_{\mu\nu}^i(x)\cdot F^{i\mu\nu}(x)
  -\frac14B_{\mu\nu}^2
  \nonumber\\
  &&+(D_\mu h(x))^{\dag}(D^\mu h(x)) 
-\lambda(\,h^{\dag}(x)\,h(x)-\mu^2)^2, \label{4.9}
\eea
where
\bea
&& F_{\mu\nu}^i (x) = \partial_\mu A_\nu^i(x)-\partial_\nu A_\mu^i(x)
       +g\,\epsilon^{ijk}A_\mu^j(x) A_\nu^k(x), \label{4.10}\\
&& B_{\mu\nu}(x)=\partial_\mu B_\nu(x)-\partial_\nu B_\mu(x),\\
&& D_\mu h(x)=\left[\partial_\mu-{i \over 2}g\,\sum_{i=1}^3 
             \tau^i\cdot  A^i_\mu(x)
             -\frac i2 g'\, B_\mu(x)
\right]h(x),\label{4.11}
\eea
with $\lambda=g_{\hbox{\rm \tiny H}}^2$ and the rescaling of 
$\mu\rightarrow \sqrt{g_{\hbox{\rm \tiny H}}}\mu$.
Except for color sector, 
Eq.(\ref{4.9}) expresses \ymh of the standard model 
with the symmetry SU(2)$_{\hbox{\rm\tiny L}}\times$U(1)$_{\hbox{\rm\tiny Y}}$
spontaneously broken to SU(1)$_{\hbox{\rm\scriptsize em}}$.
\par
Let us express the ghost and gauge fixing terms in Eqs.(\ref{2.27}) 
and (\ref{2.28}) in the case of the standard model. 
For simplicity, hereafter we abbreviate the argument $x$ 
in the respective fields.
After the same rescaling of ghost and Nakanishi-Lautrup fields such as
\bea
&&C^i({\bar C}^i)  \rightarrow g C^i({\bar C}^i),\hskip 1.0cm
   C^0({\bar C}^0)    \rightarrow g'C^0({\bar C}^0), \label{4.12}\\
&&B^i({\bar B}^i)  \rightarrow g B^i({\bar B}^i),\hskip 1.0cm
   B^0({\bar B}^0)    \rightarrow g'B^0({\bar B}^0), \label{4.12a}
\eea
we get the gauge fixing term ${\cal L}_{\hbox{\rm\tiny GF}}$ 
in Eq.(\ref{2.27}) as
\bea
  {\cal L}_{\hbox{\rm\tiny GF}}&=& 
      \sum_{i=1}^2\left\{
        \frac\alpha2{B_i}^2+
        B_i\left(\partial^\mu A_\mu^i
        +m_{\hbox{\rm\tiny W}}\phi^i\right)\right\}
         , \nonumber\\
 &&     +   \left\{\frac\alpha2B_{\hbox{\rm\tiny Z}}^2+
         B_{\hbox{\rm\tiny Z}}\left(\partial_\mu Z^\mu+
         m_{\hbox{\rm\tiny Z}}\phi^3\right)\right\}
         \nonumber\\
&&     +   \left\{ \frac\alpha2B_{\hbox{\rm\tiny A}}^2+
            B_{\hbox{\rm\tiny A}}\partial_\mu A^\mu \right\}
\label{4.13}
\eea
where $m_{\hbox{\rm\tiny{W}}}$ and $m_{\hbox{\rm\tiny{Z}}}$ are
the charged and weak neutral gauge boson masses, respectively, 
and $Z_\mu$, $A_\mu$, $B_{\hbox{\rm\tiny{Z}}}$ 
and $B_{\hbox{\rm\tiny{A}}}$ are
defined as
\bea
  &&  Z_\mu=\frac{gA_\mu^3-g'B_\mu^0}{\sqrt{g^2+{g'}^2}},\hskip1cm
      A_\mu=\frac{g'A_\mu^3+gB_\mu^0}{\sqrt{g^2+{g'}^2}}, \nonumber\\
  &&  B_{\hbox{\rm\tiny{Z}}}=\frac{gB^3-g'B^0}{\sqrt{g^2+{g'}^2}},\hskip1cm
    B_{\hbox{\rm\tiny{A}}}=\frac{{g'}B^3+gB^0}{\sqrt{g^2+{g'}^2}}.
    \label{4.14}
\eea
The parametrization of $\phi^i(i=1,2,3)$ is given in Eq.(\ref{3.16}).
By use of the equations of motion of 
the Nakanishi-Lautrup fields, Eq.(\ref{4.13}) leads to
\bea
  {\cal L}_{\hbox{\rm\tiny{GF}}}&=&
        -\sum_{i=1}^2\frac1{2\alpha}\left(\partial^\mu A_\mu^i+
        m_{\hbox{\rm\tiny{W}}}\phi^i\right)^2\nonumber\\
        &&-\frac1{2\alpha}\left(\partial_\mu Z^\mu
        +m_{\hbox{\rm\tiny{Z}}}\phi^3\right)^2
        -\frac1{2\alpha}\left(\partial_\mu A^\mu\right)^2
         , \label{4.17}
\eea
which is the gauge fixed Lagrangian
with the 't Hooft-Feynman gauge\cite{tH} when $\alpha=1$.
\par
With the same notations of ghost and anti-ghost fields as 
\bea
  &&  C_{\hbox{\rm\tiny{Z}}}=\frac{gC^3-g'C^0}{\sqrt{g^2+{g'}^2}},\hskip1cm
    C_{\hbox{\rm\tiny{A}}}=\frac{{g'}C^3+gC^0}{\sqrt{g^2+{g'}^2}}.
      \nonumber\\
  &&  {\bar C}_{\hbox{\rm\tiny{Z}}}
       =\frac{g{\bar C}^3-g'{\bar C}^0}{\sqrt{g^2+{g'}^2}},\hskip1cm
    {\bar C}_{\hbox{\rm\tiny{A}}}=\frac{{g'}{\bar C}^3
         +g{\bar C}^0}{\sqrt{g^2+{g'}^2}},
    \label{4.15}
\eea
we obtain the explicit expression of ghost terms in Eq.(\ref{2.27}) as follows:
\bea
       {\cal L}_{\hbox{\rm\tiny{GH}}}&=& 
           -i\sum_{i=1}^2\partial^\mu{\bar C}^i{\cal D}_\mu C^i
           -i\partial^\mu{\bar C_{\hbox{\rm\tiny{Z}}}}
           {\cal D}_\mu C_{\hbox{\rm\tiny{Z}}}
           -i\partial^\mu{\bar C_{\hbox{\rm\tiny{A}}}}
           {\cal D}_\mu C_{\hbox{\rm\tiny{A}}}\nonumber\\
          && +i\sum_{i=1}^2m_{\hbox{\rm\tiny{W}}}^2{\bar C}^iC^i 
              +im_{\hbox{\rm\tiny{Z}}}^2
              {\bar C}_{\hbox{\rm\tiny{Z}}}C_{\hbox{\rm\tiny{Z}}} \nonumber\\
         && +\frac{i}2\left\{{m_{\hbox{\rm\tiny{W}}}}g
         ({\bar C}^1C^1+{\bar C}^2C^2)+
         {m_{\hbox{\rm\tiny{Z}}}}\sqrt{g^2+{g'}^2}
         {\bar C}_{\hbox{\rm\tiny{Z}}}C_{\hbox{\rm\tiny{Z}}}\right\}\psi
         \nonumber\\
         && +\frac{i}2\left\{ {m_{\hbox{\rm\tiny{Z}}}}g
         {\bar C}_{\hbox{\rm\tiny{Z}}}C^2-
         {m_{\hbox{\rm\tiny{W}}}}
         \frac{(g^2-{g'}^2)}{\sqrt{g^2+{g'}^2}}
         {\bar C}^2C_{\hbox{\rm\tiny{Z}}}\right\}\phi^1 \nonumber\\
          && +\frac{i}2\left\{ 
         {m_{\hbox{\rm\tiny{W}}}}
         \frac{(g^2-{g'}^2)}{\sqrt{g^2+{g'}^2}}
         {\bar C}^1C_{\hbox{\rm\tiny{Z}}}-{m_{\hbox{\rm\tiny{Z}}}}g
         {\bar C}_{\hbox{\rm\tiny{Z}}}C^1\right\}\phi^2 \nonumber\\
         && +\frac{i}2{m_{\hbox{\rm\tiny{W}}}}g
         \left({\bar C}^2C^1-{\bar C}^1C^2\right)\phi^3\nonumber\\
        &&+ im_{\hbox{\rm\tiny{W}}}\frac{g{g'}}{\sqrt{g^2+{g'}^2}}
         \left({\bar C}^1\phi^2-{\bar C}^2\phi^1\right)C_{\hbox{\rm\tiny{A}}},
           \label{4.18}
\eea
where covariant derivatives of ghost fields are calculated through
Eqs.(\ref{4.14}) and (\ref{4.15}).
%${\cal D}_\mu C^i=\partial_\mu C^i
%+g\epsilon^{ijk}A_\mu^jC^k$ and ${\cal D}_\mu C^0=\partial_\mu C^0$.
Also in this case, the ghost fields become massive and
the new interaction terms between ghosts and
Higgs fields appear. It should be noted that ${\cal L}_{\hbox{\rm\tiny{GH}}}$
coincides with that in Eq.(\ref{3.18}) in the limit of $g'=0$
except for U(1) ghost term.\par
It should be also noted that our definition of gauge fixing condition is
connected with the prescription proposed by Kugo and Uehara \cite{KU} as
\bea
{\cal L}_{\hbox{\rm\tiny GH}+\hbox{\rm\tiny GF}}
&=&-i\partial_\theta\left[\sum_{i=1}^2{\bar C}^i\left(
\partial^\mu A^i_\mu+m_{\hbox{\rm\tiny W}}\phi^i+\frac12\alpha B^i    
\right)\right] \nonumber\\
&&-i\partial_\theta\left[{\bar C}_{\hbox{\rm\tiny Z}}
 \left(\partial^\mu Z_\mu+m_{\hbox{\rm\tiny Z}}\phi^3
+\frac12\alpha B_{\hbox{\rm\tiny Z}}\right)\right]
-i\partial_\theta\left[{\bar C}_{\hbox{\rm\tiny A}}
 \left(\partial^\mu A_\mu
+\frac12\alpha B_{\hbox{\rm\tiny A}}\right)\right]
\label{5.2}
\eea
for the standard model.
It is interesting that our definition
naturally leads to the more general gauge fixing condition.
%%%%%%%%%%%%%%%%%%%%%%%%%%%%%%%%%%%%%%%%%%%%%%%%%%%%%%%%%%%%%%%%%%%%%%%%%%%%%
\section{ Concluding remarks}
The reconstructions of the spontaneously broken gauge theories based on the
generalized differential geometry on the discrete space $M_4\times Z_N$
have consistently performed so far. Especially, the standard model is nicely
reconstructed in GDG on $M_4\times Z_2$ \cite{FSTM}
by introducing the 24-dimensional internal space where chiral fermions
are represented as 24-dimensional vectors. 
It is well understood that the Higgs boson field is a connection 
on the discrete space $Z_N$ and an unified picture of 
the ordinary gauge fields and Higgs boson field as the generalized
connection is realized. This is a common feature of the NCG approach.\par
In this paper,
BRST invariant formulation of the spontaneously broken gauge theory is 
presented in our scheme of a GDG on the
discrete space $M_4\times Z_N$.  According to the super space formulation
by Bonora and Tonin\cite{BT}, we 
introduce the Grassmann numbers $\theta$ and
${\bar \theta}$ as the arguments in super space in addition to 
$x_\mu$ in $M_4$ and $n$ in $Z_N$. 
The horizontality condition \cite{HC} on the generalized field strength 
${\cal F}(x,n,\theta,\bar{\theta})$ determines the BRST transformation of
every field including the Higgs boson field. 
By use of the generalized gauge field ${\cal A}(x,n,\theta,{\bar \theta})$
and the Nakanishi-Lautrup field ${\bar B}(x,n,\theta,{\bar \theta})$,
the gauge fixing and ghost terms are defined in Eq.(\ref{2.19})
and written explicitly in Eqs.(\ref{2.27}) and (\ref{2.28}).
Two applications to the SU(2) Higgs-Kibble model and the standard model 
show ghost fields to be massive and bring
the new interaction terms between ghosts and Higgs fields.  
This is natural because the Higgs boson is a member of the generalized
gauge field in GDG on the discrete space in the same way as ordinary gauge
fields. Especially, our BRST formulation prefers the t'Hooft-Feynman
gauge \cite{tH} as the gauge fixing condition, so in this case $\alpha=1$
is required.
\par
The Higgs mechanism necessary for the spontaneously broken gauge theory is well
understood in the generalized differential geometry on the discrete
space. In addition, the super space formalism by  Bonora and Tonin\cite{BT}
is nicely incorporated in this formulation to bring the BRST invariant
Lagrangian. It is possible to discuss the anomaly of the spontaneously 
broken gauge theory in the present formulation as a future work.
It is also an important purpose to incorporate the supersymmetry
in the present formulation. If it would be possible, this
approach would be more promising.
\vskip 0.5 cm
%%%%%%%%%%%%%%%%%%%%%%%%%%%%%%%%%%%%%%%%%%%%%%%%%%%%%%%%%%%%%%%%%%%%%%%%%%%%
%\bigskip
\begin{center}
{\bf Acknowledgement}
\end{center}
\smallskip
The author would like to
express his sincere thanks to
Professor
 H.~Kase and Professor K. Morita 
for useful suggestion and
invaluable discussions. He is grateful to all members 
at Department of Physics, Boston University for their
warm hospitality.
%%%%%%%%%%%%%%%%%%%%%%%%%%%%%%%%%%%%%%%%%%%%%%%%%%%%%%%%%%%%%%%%%%%%%%%%%%%%
%%%%%%%%%%%%%%%%%%%%%%%%%%%%%%%%%%%%%%%%%%%%%%%%%%%%%%%%%%%%%%%%%%%%%%%%%%%%
\def\jmp{J.~Math.~Phys.$\,$}
\def\pl{Phys. Lett.$\,$ }
\def\np{Nucl. Phys.$\,$}
\def\ptp{Prog. Theor. Phys.$\,$}
\def\prl{Phys. Rev. Lett.$\,$}
\def\pr{Phys. Rev. D$\,$}
\def\mp{Int. Journ. Mod. Phys.$\,$ }
%%%%%%%%%%%%%%%%%%%%%%%%%%%%%%%%%%%%%%%%%%%%%%%%%%%%%%%%%%%%%%%%%%%%%%%%%%%%

\end{document}